\documentstyle[12pt]{article}
\begin{document}
\def\lsim{\:\raisebox{-0.5ex}{$\stackrel{\textstyle<}{\sim}$}\:}
\def\gsim{\:\raisebox{-0.5ex}{$\stackrel{\textstyle>}{\sim}$}\:}
\begin{flushright}
TIFR/TH/93-23 \\
\end{flushright}
\bigskip
\bigskip
\begin{center}
\large {\bf An Effective $p_T$ Cutoff for the Isolalated Lepton Background
from Bottom Decay}\\
\bigskip
\bigskip
\large {\bf N.K. Mondal} \\
\smallskip
HECR Group\\
\smallskip
and \\
\smallskip
\large {\bf D.P. Roy}\\
\smallskip
Theoretical Physics Group\\
\smallskip
Tata Institute of Fundamental Research\\
Homi Bhabha Road, Bombay - 400 005, India\\
\bigskip
\bigskip
\underbar{\bf Abstract}\\
\end{center}
\bigskip

There is a strong correlation between the $p_T$ and isolation of the
lepton coming from $B$ decay.  Consequently the isolated lepton background
from $B$ decay goes down rapidly with increasing lepton $p_T$; and there
is a $p_T$ cutoff beyond which it effectively vanishes.  For the isolation
cut of $E^{AC}_T < 10$ GeV, appropriate for LHC, the lepton $p_T$ cutoff
is 80 GeV.  This can be exploited to effectively eliminate the $B$
background from the like sign dilepton channel apropriate for Majorana
particle searches, as well as the unlike sign dilepton and the single
lepton channels appropriate for the top quark search.  We illustrate this
with a detailed analysis of the $B$ background in these channels along
with the signals at LHC energy using both parton level MC and ISAJET
programs.

\newpage

\noindent \underbar{\bf I. Introduction}\\

\bigskip
It was first pointed out a decade back [1], that the prompt lepton
background from bottom and charm decay can be largely suppressed by the
isolation criterion without affecting a top quark signal.  Since then the
lepton isolation criterion has been extensively used in the search of top
quark as well as several other heavy particles, particularly at hadron
colliders.  The small mass of the bottom and charm relative to the lepton
$p_T$ implies that the lepton emerges together with the decay hadrons
within a narrow cone, and hence can be suppressed by a suitable isolation
criterion, i.e.
$$
E^{AC}_T < E_0
\eqno (1)
$$
where $E^{AC}_T$ is the transverse energy accompanying the lepton within a
narrow cone.  The half-angle of the cone is usually taken to be about 0.4
radians, or the angular radius
$$
\Delta R = \sqrt{\Delta \phi^2 + \Delta \eta^2} < 0.4
\eqno (2)
$$
where $\Delta \phi$ and $\Delta\eta$ are the azimuthal angle and
pseudo-repidity of the particles relative to the lepton.  Only a small
fraction of the bottom and charm decay leptons, where the accompanying
decay hadrons are very soft $(E_T < E_0)$, survive this isolation cut.

Subsequently it was pointed out in [2] that the above lepton isolation cut
becomes increasingly more effective in suppressing the bottom and charm
background with increasing $p_T$ of the lepton.  It was further suggested
that for a given isolation cut (1), the lepton $p_T$ has an approximate
upper limit given by
$$
p^\ell_T \lsim E^{AC}_T \left({m^2_B - m^2_C \over m^2_C}\right)
\eqno (3)
$$
for bottom particle decay
$$
B \rightarrow \ell \nu C
\eqno (4)
$$
and an analogous limit for the charm particle decay, where $\ell$ stands
for either $e$ or $\mu$.  Since then the correlation between the lepton
$p_T$ and isolation has been used [3] for suppressing the bottom and charm
background to a top quark signal.  However, the background suppression
potential of this correlation and in particular the lepton $p_T$ cutoff
(3) implied by it have not been fully explored.  Nor has its application
been extended beyond the top search programme so far.  The present work is
devoted to a systematic study of these issues.

To be specfic we shall concentrate on the leptons coming from bottom
decay, i.e. eq. (4).  We favour this proces for two reasons.  Firstly, the
bottom decay leptons constitute a moe serious background to top signal
compared to charm.  Secondly, and more importantly, the bottom decay
leptons constitute the most serious standard model background to the like
sign dilepton (LSD) channel, which is the canonical channel for the search
of heavy Majorana particles (e.g. Majorana neutrinos of $L-R$ models and
the neutralinos of SUSY models).  Indeed the elimination of the leptonic
background from bottom decay is more important in this context than that
of top quark search, since the latter has a large signal as well as a
serious background from $W$ decay.

We shall start section II with a derivation of the approximate $p_T$
cutoff formula (3) for isolated leptons coming from bottom decay, which
has not been given in the literature so far.  The effects neglected in the
derivation of this formula will be identified.  In section III we shall
make a quantitative estimate of the leptonic background from bottom decay
in the LSD channel for a hadron collider and study the correlation between
the isolation and $p_T$ of the lepton using a parton level Monte Carlo [4]
as well as the ISAJET program [5].  In particular the effects neglected in
derieving the approximate $p_T^\ell$ cutoff formula (3) and the resulting
spill over of the background beyond this cutoff will be explored in
detail.  Thus we shall see that there is an effective $p_T$ cutoff for the
isolated lepton background from bottom decay, which is not too far from
the approximate value given by (3).  Then we shall consider some specific
examples of Majorana particle signals in the isolated LSD channel.  We
shall see that the effective $p_T$ cutoff practically eliminates the
background while retaining a large fraction of these signals.  In section
IV we shall present a similar analysis of the isolated lepton background
from bottom decay in the context of top quark search.  Our results will be
summarised in section V.  To be specific we shall work with the LHC
collider energy of
$$
\sqrt{s} = 1.6 ~{\rm TeV}.
$$
However the result should be equally valid for the SSC.
\bigskip
\bigskip

\noindent \underbar{\bf II. Correlation Between Lepton} $\underline{p_T}$
\underbar{\bf and Isolation}: \\

\bigskip
\nobreak
Fig. 1 shows a $B$ particle emerging at an angle $\theta$ relative to the
beam axis with lab momentum $p_B$.  Let us derieve the approximate formula
(3) for the maximum value of the decay lepton $p_T$ for a given
accompanying $E_T$.  We assume
$$
p^\ell_T \gg m_B
\eqno (5)
$$
implying $p^\ell$ and $p_B \gg m_B$, i.e. the lab momenta of all the decay
particles from (4) are collimated in the direction of $p_B$.  Thus the
maximum value of $p_T^\ell$ relative to $E^C_T$ corresponds to that of
$p^\ell$ relative to $E^C$ (or $p^C$).  This corresponds to the decay
configuration in the $B$ rest frame shown in Fig. 1, where $\ell$ and $C$
are emitted along and opposite to the direction of $p_B$ with the maximum
allowed momentum $q$ together with a soft $\nu$.  The corresponding
rapidity difference is
$$
Y^\star_\ell - Y^\star_C = \ell n\left({q + E^\star_\ell \over
m_\ell}\bigg/{E^\star_C - q \over m_C}\right) = \ell n {m^2_B - m^2_C
\over m_c m_\ell}
\eqno (6)
$$
neglecting $m_\ell$ with respect to $q$ and $m_C$.  The corresponding
rapidity difference in the lab frame is
$$
Y_\ell - Y_C = \ell n\left({2p^\ell \over m_\ell}\bigg/{E^C + p^C \over
m_C}\right);
\eqno (7)
$$
and since the rapidity difference is invariant under longitudinal boost,
we have
$$
E^C + p^C = 2p^\ell \cdot m^2_C/(m^2_B - m^2_C).
\eqno (8)
$$
Correspondingly
$$
E^C - p^C = (m^2_B - m^2_C)/2p^\ell
\eqno (9)
$$
i.e.
$$
E^C = p^\ell {m^2_C \over m^2_B - m^2_C} + {m^2_B - m^2_C \over 4p^\ell}.
\eqno (10)
$$
Since this is the maximum value of $p^\ell$ relative to $E^C$, we have
$$
E^C \geq p^\ell {m^2_C \over m^2_B - m^2_C}
\eqno (11)
$$
which immediately leads to (3).  One sees that the maximum value of
$p^\ell$ relative to $p^C$ (or $E^C$) is independent of $p_B$; it is
determined by the ratio $m_C/m_B$.  Note that one can identity $E^C_T$
with the accompanying $E_T$, provided
$$
E^{AC}_T \gg m_C,
\eqno (12)
$$
which is automatically guaranteed by (11) and (5).  However, the question
is how well is it satisfied in the real experimental situations?  The UA1
[6] and CDF [7] experiments have used $E^{AC}_T$ cuts of 2 GeV and 5 GeV
respectively, while the canonical value for LHC is taken to be [3]
$$
E^{AC}_T < 10 ~{\rm GeV}.
\eqno (13)
$$
Thus we expect the lepton $p_T$ cutoff (3) to be good for the LHC and to a
lesser extent for the CDF data as well, but not for the UA1.

Substituting $m_B = 5.28$ GeV and $m_C = m_D (m^\star_D) = 1.87 (2.01)$
GeV [8] in (3) gives
$$
p^\ell_T \lsim 7 E^{AC}_T (6 E^{AC}_T).
\eqno (14)
$$
Thus for the canonical $E^{AC}_T$ cut of 10 GeV at LHC, we get
$$
p^\ell_T \lsim 70 ~{\rm GeV} ~(60 ~{\rm GeV}),
\eqno (15)
$$
corresponding to $B \rightarrow \ell \nu D ~(D^\star)$.  Note that the
stronger limit of 60 GeV corresponds to the dominant decay mode of $B$
[8].

It should be emphasised that the origin of the above $p_T$ cutoff for the
isolated lepton background from $B$ decay is distinct from that of the
standard isolation $(E^{AC}_T)$ cut.  While the isolation cut originated
from the small value of the ratio $m_B/p^\ell_T$, the $p^\ell_T$ cutoff
arises from the sizeable value of the ratio $m_C/m_B$ which is a distinct
physical condition.  The former ensures a small opening angle between
$\ell$ and $C$, while the latter guarantees a minimum $E^C_T$ for a given
$p^\ell_T$.  Indeed the strong correlation between the lepton $p_T$ and
isolation arises almost entirely due to the latter condition.  We have
checked this through an explicit MC calculation; the effect of the
isolation cut (13) becomes practically independent of $p^\ell_T$ as we
make $m_C \ll m_B$.

A $p_T$ cutoff of 60 -- 70 GeV for the isolated lepton background from $B$
decay is evidently a powerful result.  But it is only an approximate one,
since it has not taken account of the following effects.  1) For the
semileptonic decay of $D$
$$
D \rightarrow \ell \nu K (K^\star),
\eqno (16)
$$
occuring with an average branching fraction of 30\%, a part of $D$ energy
is lost to the neutrino.  This would cause a reduction in the visible
$E^{AC}_T$ or equivalently a spill over of $p^\ell_T$ beyond the limit
(15).  2) While the bulk of the energy of the $b$ quark jet is carried by
the $B$ hadron, a part of the remainder is expected to come out within the
cone (2).  This would enhance the $E^{AC}_T$ or equivalently strengthen
the $p^\ell_T$ limit further.  We shall use the Peterson fragmentation
model [9] for $b$ quark in to $B$ hadron
$$
P^B_b (Z) = N\Big/Z\left(1 - {1 \over Z} - {\epsilon \over 1 - Z}\right)^2,
\eqno (17)
$$
which denotes the probability of $B$ hadron carrying an energy-momentum
fraction $Z$.  We shall take $\epsilon = 0.006$, corresponding to $\langle
Z\rangle_B = 0.83$ [7].  What fraction of the remainder $(1-Z)$ comes
within the cone (2), can be estimated only via a detailed fragmentation
model.  This is done in the ISAJET program.  To estimate the size of this
effect, however, we shall also present the parton level MC results for the
two extreme cases where all or none of the remaining $b$-jet energy is
added to $E^{AC}_T$.  3) Finally the resolution smearing of $E^{AC}_T$
will cause a small spill over of the lepton $p_T$ beyond the limit (15).
This will be taken in to account by incorporating a typical hadron energy
resolution
$$
\Delta E/E = 0.5/\sqrt{E}
\eqno (18)
$$
in GeV units.  The effect of lepton energy resolution is negligible.

The most important mechanisms of $b$ quark production at LHC are the
direct $b\bar b$ production via gluon-gluon fusion
$$
gg \rightarrow b\bar b
\eqno (19a)
$$
$$
{}~~~~~~\rightarrow b\bar b g
\eqno (19b)
$$
and via top quark decay
$$
gg \rightarrow t\bar t(\rightarrow \bar b).
\eqno (20)
$$
For the large $p_T$ region of our interest the dominant contribution comes
from the next to leading order QCD process (19b) in the first case while
the leading order contribution is adequate in the second case.  The
relevant formulae for the next to leading order QCD process are given in
[10].  Those for the leading order process can be found in [11] along with
the formulae for the semileptonic decay of $B$ and $D$ hadrons.  We shall
use the free quark decay formulae for $t$.

For top quark search, of course, one needs to consider only the bottom
quark background coming from (19).  However, both (19) and (20) contribute
to the background in the like sign dilepton channel.  While the latter
contributes in full strength, the former contributes via $B\bar B$ mixing.
One has to multiply the dilepton cross-section coming from (19) by the
mixing factor
$$
\chi (1 - \chi), ~~~\chi \simeq 0.15
\eqno (21)
$$
to obtain the cross-section for the $\ell^+\ell^+$ channel [12].  This is
the dominant contribution at low $p_T$, while the contribution from (20)
dominates in the large $p_T$ region of our interest.  The reason is
two-fold.  Firstly, only one of the leptons come from $b$ in the latter
case so that one has to pay the price of isolation once.  Secondly the $b$
has a comparatively larger $p_T$ in this case resulting in a harder $p_T$
spectrum for the decay lepton.  This has also a big practical advantage in
analysing the correlation between the lepton $p_T$ and isolation -- i.e.
the analysis can be carried over a large range of $p_T$ with a manageable
numer of MC events.  Therefore we start the analysis below with this
process.

\bigskip

\noindent \underbar{\bf III. $B$ Background in the Like Sign Dilepton
Channel} \\

\underbar{\bf (Search for Majorana Particles)}\\

\bigskip
\nobreak
Fig. 2 shows the like sign dilepton cross-section $\sigma_{\ell^+\ell^+}$
from (20) against the $p_T$ of the 2nd (softer) lepton, which comes from
$B$ decay.  The upper histogram and the solid curve show this lepton $p_T$
distribution without any isolation cut for the ISAJET and the parton level
MC programs respectively.  The lower histogram and the lower curves refer
to the corresponding distributions with the isolation cut $E^{AC}_T < 10$
GeV -- the different curves refer to different stages of approximation in
evaluting $E^{AC}_T$ as we shall see below.  Comparing first the two
histograms we see a very strong correlation between the lepton $p_T$ and
isolation.  The isolation cut reduces this leptonic background from $B$
decay by only a factor of $\sim 10$ at $p_T = 20$ GeV.  But the reduction
factor increases to $\sim 100$ at $p_T = 50$ GeV and 400 at $p_T = 60$
GeV.  The tail extending beyond this point arises from the loss of visible
$E^{AC}_T$ due to the $D$ semileptonic decay (16).  It has the effect of
increasing the lepton $p_T$ cutoff from 60 to 80 GeV.  There is only 1 MC
event lying beyond 80 GeV (see Fig. 3) out of a total of 351383.  This 1
event corresponds to 1/2 event for the projected luminosity of 10
fb$^{-1}$ at LHC.  Since the minimum size of any viable signal has to be
$\geq 10$ events, one can safely neglect the above background compared to
this.  Thus one has an effective cutoff of
$$
p^\ell_T \leq 80~{\rm GeV}
\eqno (22)
$$
for $B$ decay leptons with $E^{AC}_T < 10$ GeV.

The dot-dashed (dotted) line shows the parton level MC distribution with
(without) energy resolution for the simplest model, which does not take
into account the reduction of $E^{AC}_T$ due to the semileptonic decay of
$D$ or its enhancement due to the fragments of the $b$-quark jet.  In
other words, this is the model which lead to the approximate $p_T$ limit
(3).  One should note that this model prediction is close to the ISAJET
distribution all the way up to 60 GeV.  The long dashed line includes the
effect of $D$ semileptonic decay.  The short dashed line shows the maximal
effect of the fragments of the $b$ quark jet -- i.e. it corresponds to all
the fragments emerging within the $E^{AC}_T$ cone (2).  As remarked
before, one can not say what fraction of these fragments are contained
within this cone without assuming a detailed fragmentation model, which is
outside the scope of the parton level MC program.  None the less one can
set limits corresponding to all and none of these fragments (other than
the $B$ hadron) coming within the $E^{AC}_T$ cone.  These are given by the
short and long dashed lines respectively.  Note that the ISAJET
distribution lies in between these limits.  One should also note that even
the upper limit of this background, represented by the long dashed lines,
has an effective $p_T$ cutoff at 80 GeV.  This is an important result.
For the fragmentation model assumed to fix what fraction of the $b$-quark
jet fragments are contained within the $E^{AC}_T$ cone is the least tested
part of the ISAJET program.  Finally the crosses represent the
contribution from (19) to the isolated $\ell^+\ell^+$ channel via $B\bar
B$ mixing.  As remarked before, this contribution is larger than that from
(20) at small $p_T$ but falls below the latter in the large $p_T$ region
of our interest.

Fig. 3 shows the ISAJET results for the $B$ background along with some
examples of Majorana particle signal.  The $B$ background is shown with
$E^{AC}_T$ cuts of 10 and 5 GeV as well as without any isolation
$(E^{AC}_T)$ cut.  The signals are shown with $E^{AC}_T$ cut of 10 GeV,
but the depence on $E^{AC}_T$ is marginal; the bulk of the signal events
come with isolated leptons.  We see that the isolated lepton background
from $B$ decay can be practically eliminated with a $p_T$ cut of 80 GeV,
while retaining substancial parts of the signals.  The dashed lines
represent the right-handed Majorana neutrino signal for $M_{NR} = 200$ and
1000 GeV with a right-handed $W$ boson mass of 3000 GeV [13].  The signal
size increases with decreasing $W_R$ mass and vice versa.  This
prescription for background elimination is equally valid for other types
of Majorana neutrino signals not shown here [14].  The solid lines
represent the gluino pair production signal in the $R$-parity breaking
SUSY model for $M_{\tilde g} = 600$ and 1000 GeV.  Each gluino decays into
a photino $\tilde\gamma$ either directly or cascading [15] via the
intermediate gauginos $(\tilde Z,\tilde W)$.  The leptonic decay of
$\tilde\gamma$ in the $R$-parity breaking SUSY model leads to the LSD
signal.  The lines represent conservative estimates of this signal coming
from the direct decay of each $\tilde g$ into $\tilde \gamma$ followed by
the leptonic decay of $\tilde\gamma$ [16].  The overall branching fraction
for this process is taken to be 2\%, corresponding to a 14\% branching
fraction for the direct decay of each $\tilde g$ [15,17].  Further LSD
events are expected to come from the cascade decay process.  Indeed these
events are expected to provide a viable LSD signal for gluino even without
$\tilde\gamma$ decay -- i.e. in the $R$ conserving SUSY model [17].  The
above prescription for background elimination would be equally relevant
for this signal as well.

Fig. 4 shows the distribution of this LSD background in the complimentary
variable $E^{AC}_T$, for 4 values of the lepton $p_T$ cut -- 20,40,60 and
80 GeV.  One can clearly see the depletion of the low $E^{AC}_T$ (Isolated
lepton) background with the increasing $p_T$ cut of the 2nd ($B$ decay)
lepton.  Thus with a $p_T$ cut of 80 GeV, there are no events with
$E^{AC}_T < 8$ GeV and only 2 events in the $E^{AC}_T = 8 - 16$ GeV region
(corresponding to 1 event at LHC).  The size of isolated LSD signals for
the above mentioned Majorana particles are also shown for a lepton $p_T$
cut of 80 GeV, spread over an $E^{AC}_T$ bin of 0 -- 4 GeV.  One can
clearly separate the signals from the background with a $E^{AC}_T$
resolution of $\sim 8$ GeV, thanks to the large lepton $p_T$ cut.

\bigskip

\noindent \underbar{\bf IV. $B$ Background in the Single Lepton and Unlike}\\

\nobreak
\underbar{\bf Sign Dilepton Channels (Top Quark Search)}\\

\bigskip
\nobreak
We shall finally consider the leptonic background from $B$ decay in the
context of top quark search.  Although parts of this issue have been
discussed earlier [3], we feel it will be worth while to cover it for the
sake of completeness.  The relevant production process for the $B$
background in this case is the QCD process (19).  In the large $p_T$
region of our interest the dominant contribution comes from the next to
leading order process (19b).  The parton level MC results have been
computed with (19b) while the ISAJET results include the lowest order
process (19a) as well.  But the ISAJET result has been normalised to the
parton level MC cross-section as in the previous case.

Fig. 5 shows the lepton $p_T$ distribution of the $B$ background in the
single lepton channel.  The upper histogram and solid curve are the ISAJET
and parton level MC results without any isolation cut.  The lower
histogram represents the ISAJET result for the isolation cut of $E^{AC}_T
< 10$ GeV.  The dot-dashed (long dashed) lines are the corresponding
parton level MC results without (with) the effect of $D$ semileptonic
decay.  Comparison of the two histograms shows a strong correlation
between the lepton $p_T$ and isolation as in the previous case.  Thus the
suppression of the $B$ background due to the isolation cut is seen to
increase from a factor of only $\sim 3$ at $p^\ell_T = 20$ GeV to $\sim
40$ at 50 GeV [18].  Unfortunately one can not go further in $p^\ell_T$ as
one runs out of MC statistics.  This is mainly due to the relatively soft
$p^\ell_T$ spectrum from (19) compared to (20), as remarked before.  There
are no isolated lepton events left beyond $p^\ell_T = 72$ GeV out of a
total sample of over 2.2 million MC events.  Turning to the parton level
MC distributions we see that the simplest model prediction (dot-dashed
line) shows a lepton $p_T$ cutoff of $\sim 60$ GeV as expected from
(14,15).  Taking account of the $D$ semileptonic decay (long dashed line)
increases the effective $p_T$ cutoff to 80 GeV.  It should be remembered
of course that the long dashed line corresponds to an upper limit of the
isolated lepton background as it does not include any contribution to
$E^{AC}_T$ from the fragments of the $b$ quark jet (other than the $B$
hadron).  Finally, the short dashed lines show the expected isolated
lepton signals from a top quark of mass = 150 and 200 GeV.  A lepton $p_T$
cut of 80 GeV is seen to effectively eliminate the $B$ background while
retaining a significant part of the top signal.  Unlike the LSD channel,
however, the background is not reduced to $\lsim 1$ event for the typical
LHC luminosity of 10 fb$^{-1}$.  Nor is it possible to determine at what
value of lepton $p_T$ cut will this be achieved, with the present level of
MC statistics.  But we feel this is not a worth while exercise, since (i)
the signal is more than an order of magnitude larger and (ii) there is an
equally large isolated lepton background from $W$ plus QCD jets.

Fig. 6 shows the complimentary distributions of the $B$ background in
$E^{AC}_T$ for the lepton $p_T$ cuts of 20, 40 and 50 GeV.  It also shows
the isolated lepton signal for a 150 GeV top, spread over a $E^{AC}_T$ bin
of 4 GeV, for the lepton $p_T$ cuts of 20 and 50 GeV.  Increasing the
lepton $p_T$ cut from 20 to 50 GeV is clearly seen to increase the signal
to background ratio as well as the kinematic separation between the two.

Finally, Fig. 7 shows the $b\bar b$ background and the top quark signal
for the unlike sign $e\mu$ channel against the $p_T$ of the 2nd (softer)
lepton.  This is known to be the cleanest channel for top quark search
since it does not have the background from $W$ plus QCD jets; and the $WW$
background is more than an order of magnitude smaller [3].  The $b\bar b$
background is seen to go down very rapidly with increasing lepton $p_T$,
thanks to the isolation cut on both the leptons.  Thus increasing the
lepton $p_T$ cut from 20 to 50 GeV reduces the background by 4 orders of
magnitude and increases the signal to background ratio from less than 1/10
to more than 100.  One may also note that increasing the $p_T$ cut to 80
GeV would reduce the background to $< 1$ event for the LHC luminosity of
10 fb$^{-1}$.  Thus one has an effective lepton $p_T$ cutoff of 80 GeV for
the bottom background in the isolated $e^\pm \mu^\mp$ channel as well.  Of
course a complete elimination of this background is not crucial for the
top search program, given $M_t \lsim 200$ GeV.

\bigskip

\noindent \underbar{\bf V. Summary}\\

\bigskip
\nobreak
We have studied the correlation between the $p_T$ and isolation of the
lepton background from $B$ decay $(B \rightarrow C\ell\nu)$ at LHC energy,
using a parton level MC as well as ISAJET program.  The main results of
our analysis are as follows.

\begin{enumerate}

\item[{1)}] There is a strong collelation between the lepton $p_T$ and
isolation i.e. the accompanying $E_T$.  Consequently the isolation
$(E^{AC}_T)$ cut becomes increasingly more powerful with increasing $p_T$
of the lepton.  Moreover there is a lepton $p_T$ cutoff $\simeq (6-7)
E^{AC}_T$.  The physical origin of the correlation and the resulting
lepton $p_T$ cutoff is distinct
from that of the isolation cut.  The former arise from the sizeable
value of the
ratio $m_C/m_B$ while the latter arose from the small value of
$m_B/p^\ell_T$.

\item[{2)}] The typical isolation cut of $E^{AC}_T < 10$ GeV at LHC
implies a $p_T$ cutoff $\sim 60 - 70$ GeV for isolated leptons coming from
$B$ decay.  After accounting for the reduction of visible $E^{AC}_T$ due
to the semileptonic decay of $C$ and energy resolution, one gets an
effective $p_T$ cutoff = 80 GeV.  Thus the isolated lepton background from
$B$ decay can be practically eliminated at LHC with a $p_T$ cut of 80 GeV.

\item[{3)}] This is important for several cases of new particle search for
which the dominant background comes from $B$ decay leptons.  The 2 most
important examples are (i) the like sign dilepton channel for Majorana
particles, and (ii) the unlike sign dilepton as well as the single lepton
channels for the top quark.

\item[{4)}] The $B$ background for the like sign dilepton channel is
analysed in detail along with two interesting examples of Majorana
particle signals (i.e. the heavy right handed neutrino of $L-R$ model and
the gluino of $R$-parity breaking SUSY model).  A lepton $p_T$ cut of 80
GeV is shown to reduce the background to $< 1$ event for the LHC
luminosity of 10 fb$^{-1}$, while retaining substantial parts of the
signals.  Consequently the search for these Majorana particles can be
carried upto the TeV mass range in a practically background free
environment at LHC.

\item[{5)}] The single lepton and the unlike sign dilepton $(e^\pm
\mu^\mp)$ channels for top quark search are also covered for the sake of
completeness, although parts of this have been studied before.  The
correlation between the lepton $p_T$ and isolation is particularly strong
for the $B$ background in the $e\mu$ channel, which is the cleanest
channel for top quark search.  Thus increasing the lepton $p_T$ cut from
20 to 50 GeV increases the signal to background ratio from less than 1/10
to more than 100.  Increasing it further to 80 GeV would again reduce the
background to $< 1$ event for the LHC luminosity of 10 fb$^{-1}$ ; but a
complete elimination of this background is not essential for top search,
given $M_t \lsim 200$ GeV.

\end{enumerate}
\bigskip

It is a pleasure to thank Tariq Aziz, Debajyoti Choudhury and Rohini
Godbole for discussions.

\newpage

\begin{center}
\underbar{\bf References}\\
\end{center}

\begin{enumerate}

\item{} R.M. Godbole, S. Pakvasa and D.P. Roy, Phys. Rev. Lett. 50, 1539
(1983); see also V. Barger, A.D. Martin and R.J.N. Phillips, Phys. Rev.
D28, 145 (1983).

\item{} D.P. Roy, Phys. Lett. B196, 395 (1987).

\item{} F. Cavanna, D. Denegri and T. Rodrigo, Proc. LHC Workshop, Vol.
II, 329, CERN 90-10 (1990).

\item{} The parton level MC calculations are performed with the QCD
parametrisation of M. Gluck, F. Hoffmann and E. Reya, Z. Phys. C13, 119
(1982).

\item{} F.E. Paige and S.D. Protopopescu, ISAJET Program, BNL-38034
(1986).

\item{} UA1 Collaboration: C. Albazar et al., Z. Phys. C37, 505 (1988).

\item{} CDF Collaboration: F. Abe et al., Phys. Rev. D45, 3921 (1992).

\item{} Particle Data Group, Phys. Rev. D45, S1 (1992).

\item{} C. Peterson, D. Schlatter, I. Schmitt and P. Zerwas, Phys. Rev.
D27, 105 (1983).

\item{} R.K. Ellis and J.C. Sexton, Nucl. Phys. B282, 642 (1987).

\item{} D.P. Roy, Z. Phys. C21, 333 (1984).

\item{} T. Aziz, TIFR internal note, November 1990 (Unpublished); Plenary
Talks by P. Roudeau and M.V. Danilov, Proc. Intl. Lepton-Photon Symp. and
Europhysics Conf. on High Energy Physics, Geneva, 1991, Vol. 2 (World
Scientific, 1992).

\item{} A. Datta, M. Guchait and D.P. Roy, Phys. Rev. D47, 961 (1993).

\item{} D.A. Dicus and P. Roy, Phys. Rev. D44, 1593 (1991); D. Choudhury,
R.M. Godbole and P. Roy, Phys. Lett. B (to be published).

\item{} H. Baer, V. Barger, D. Karatas and X. Tata, Phys. Rev. D36, 96
(1987).

\item{} D.P. Roy, Phys. Lett. 283B, 270 (1992).

\item{} R.M. Barnett, J.F. Gunion and H.E. Haber, Snow mass: DPF Summer
Study 1988, 230 (1988).

\item{} Note that these suppression factors are 2-3 times smaller than the
corresponding factors of Fig. 2.  This may be largely due to the effective
$p_T$ for a given bin of Fig. 5 being smaller than the corresponding bin
of Fig. 2.

\end{enumerate}
\newpage

\begin{center}
\underbar{\bf Figure Captions}\\
\end{center}

\begin{enumerate}

\item[{Fig. 1.}] The decay configuration of $B \rightarrow \ell\nu C$ in
the $B$ rest frame, corresponding to the maximum lab momentum of lepton
$p^\ell$ relative to $p^C$ (or $E^C$).

\item[{Fig. 2.}] The like sign dilepton cross-section from (20) shown
against the $p_T$ of the 2nd ($B$ decay) lepton using ISAJET (histograms)
and parton level MC program (curves).  For the $E^{AC}_T < 10$ GeV cut,
the different curves represent the simplest model with $E^{AC}_T = E^C_T$
(dotted), followed by the inclusion of energy resolution (dot-dashed) and
$D$ semileptonic decay (long dashed).  The short dashed line corresponds
to the maximal enhancement of $E^{AC}_T$ by the fragments of the $b$-quark
jet.  The contribution from (19) is shown by the crosses.

\item[{Fig. 3.}] The ISAJET prediction for the $B$ background in the like
sign dilepton channel compared with the signals for right-handed Majorana
neutrino (dashed) and gluino (solid) production.  The signals are
calculated with $E^{AC}_T < 10$ GeV cut.

\item[{Fig. 4.}] The ISAJET prediction for the $B$ background in the like
sign dilepton channel shown as a function of $E^{AC}_T$ for different
lepton $p_T$ cuts.  The right-handed Majorana neutrino and gluino signals
are shown for the lepton $p_T$ cut of 80 GeV.

\item[{Fig. 5.}] The $B$ background to the single lepton cross-section
from (19) shown as a function of the lepton $p_T$ along with the top quark
signal (short dashed lines).  The convention for the background curves are
as in Fig. 2.

\item[{Fig. 6.}] The ISAJET prediction for the $B$ background in the
single lepton channel shown as a function of $E^{AC}_T$ for different
lepton $p_T$ cuts.  The signal for a 150 GeV top quark is shown for the
lepton $p_T$ cuts of 20 and 50 GeV.

\item[{Fig. 7.}] The $B$ background to the unlike sign $e\mu$ channel from
(19) shown against the softer lepton $p_T$ along with the top quark signal.

\end{enumerate}
\end{document}